\documentclass[sigconf, nonacm]{acmart}
\usepackage{multirow} 
\usepackage{subfig}
\usepackage{cleveref}
\newfloat{floatquote}{tbp}{loq}
\floatname{floatquote}{Supplementary Text}
\crefname{floatquote}{quote}{quotes}
\Crefname{floatquote}{Supplementary Text}{Quotes}
\AtBeginDocument{%
  }

\setcopyright{acmlicensed}
\copyrightyear{2018}
\acmYear{2018}
\acmDOI{XXXXXXX.XXXXXXX}

\acmConference[Conference acronym 'XX]{Make sure to enter the correct
  conference title from your rights confirmation emai}{June 03--05,
  2018}{Woodstock, NY}
\acmISBN{978-1-4503-XXXX-X/18/06}

\begin{document}

\title{"A Great Start, But...": Evaluating LLM-Generated Mind Maps for Information Mapping in Video-Based Design}

\author{Tianhao He}
% \orcid{0000-0002-8486-3940}
\affiliation{%
  \institution{Delft University of Technology}
  \streetaddress{Landbergstraat 15}
  \city{Delft}
  \postcode{2628 CE}
  \country{The Netherlands}}
\email{t.he-1@tudelft.nl}

\author{Karthi Saravanan}
\affiliation{%
  \institution{Delft University of Technology}
  \streetaddress{Landbergstraat 15}
  \city{Delft}
  \postcode{2628 CE}
  \country{The Netherlands}}
% to Karthi: your affiliation goes here
\email{karthi.s.official@gmail.com}

\author{Evangelos Niforatos}
% \orcid{0000-0002-0484-4214}
\affiliation{%
  \institution{Delft University of Technology}
  \streetaddress{Landbergstraat 15}
  \city{Delft}
  \country{Netherlands}}
\email{e.niforatos@tudelft.nl}

\author{Gerd Kortuem}
\affiliation{%
  \institution{Delft University of Technology}
  \streetaddress{Landbergstraat 15}
  \city{Delft}
  \postcode{2628 CE}
  \country{The Netherlands}}
\email{g.w.kortuem@tudelft.nl}
\renewcommand{\shortauthors}{He et al.}

\begin{abstract}
Extracting concepts and understanding relationships from videos is essential in Video-Based Design (VBD), where videos serve as a primary medium for exploration but require significant effort in managing meta-information. Mind maps, with their ability to visually organize complex data, offer a promising approach for structuring and analysing video content. Recent advancements in Large Language Models (LLMs) provide new opportunities for meta-information processing and visual understanding in VBD, yet their application remains underexplored. This study recruited 28 VBD practitioners to investigate the use of prompt-tuned LLMs for generating mind maps from ethnographic videos. Comparing LLM-generated mind maps with those created by professional designers, we evaluated rated scores, design effectiveness, and user experience across two contexts. Findings reveal that LLMs effectively capture central concepts but struggle with hierarchical organization and contextual grounding. We discuss trust, customization, and workflow integration as key factors to guide future research on LLM-supported information mapping in VBD.
\end{abstract}

\begin{CCSXML}
<ccs2012>
   <concept>
       <concept_id>10003120.10003121.10011748</concept_id>
       <concept_desc>Human-centered computing~Empirical studies in HCI</concept_desc>
       <concept_significance>500</concept_significance>
       </concept>
   <concept>
       <concept_id>10010147.10010178.10010199.10010200</concept_id>
       <concept_desc>Computing methodologies~Planning for deterministic actions</concept_desc>
       <concept_significance>500</concept_significance>
       </concept>
 </ccs2012>
\end{CCSXML}

\ccsdesc[500]{Human-centered computing~Empirical studies in HCI}
\ccsdesc[500]{Computing methodologies~Planning for deterministic actions}

\keywords{Information Mapping, Large Language Model, Video-based Design, Designer-AI Collaboration}

\begin{teaserfigure}
  \centering
      \subfloat[\centering A screenshot of the study platform with a video player (left) and a mind map editing panel (right) for visualizing and modifying mind maps.]{{\includegraphics[width=0.55\linewidth]{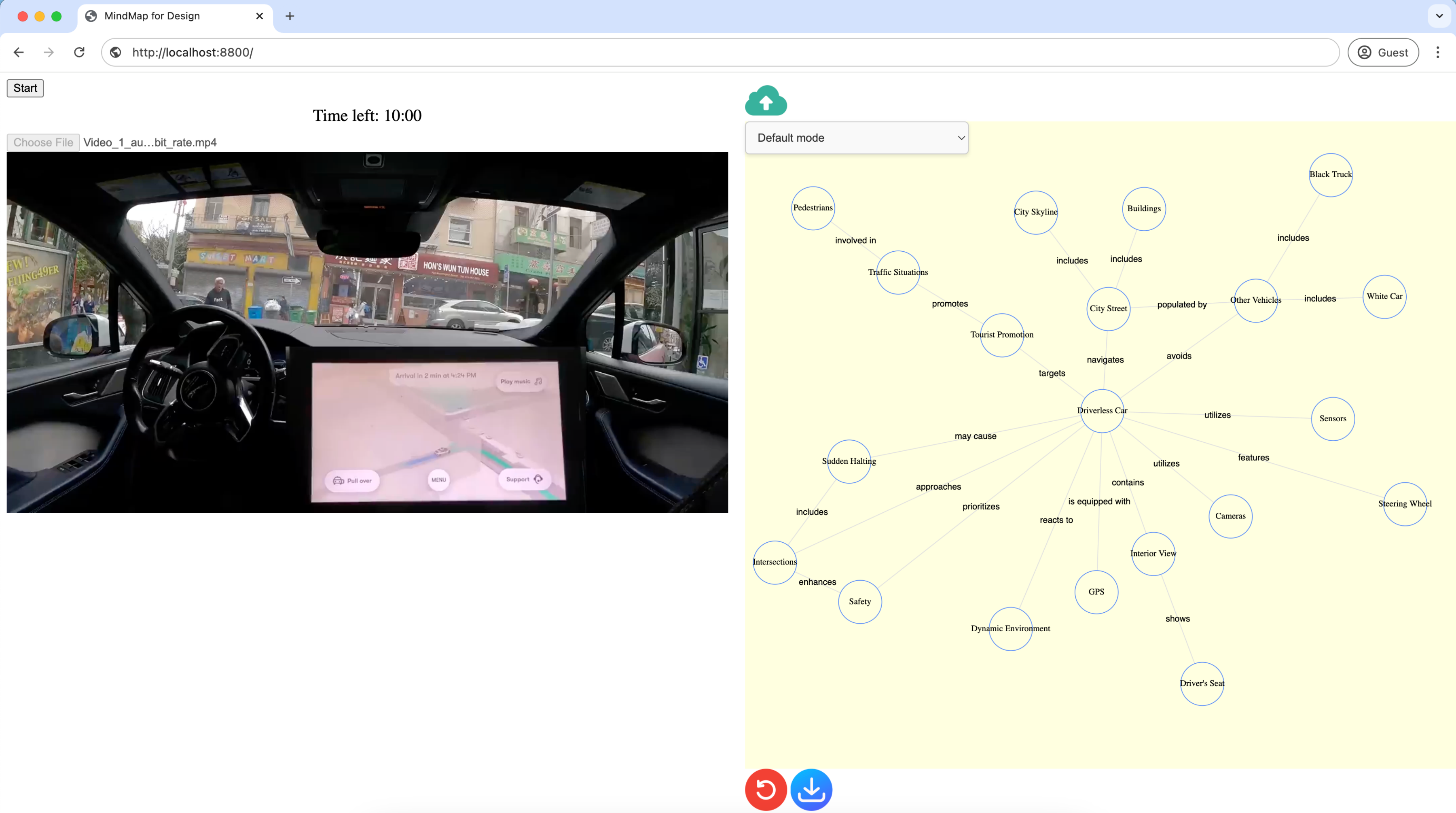}}
    \label{fig:platform}}%
    %\qquad    
      \subfloat[\centering Participant viewing videos during study]{{\includegraphics[width=0.44\linewidth]{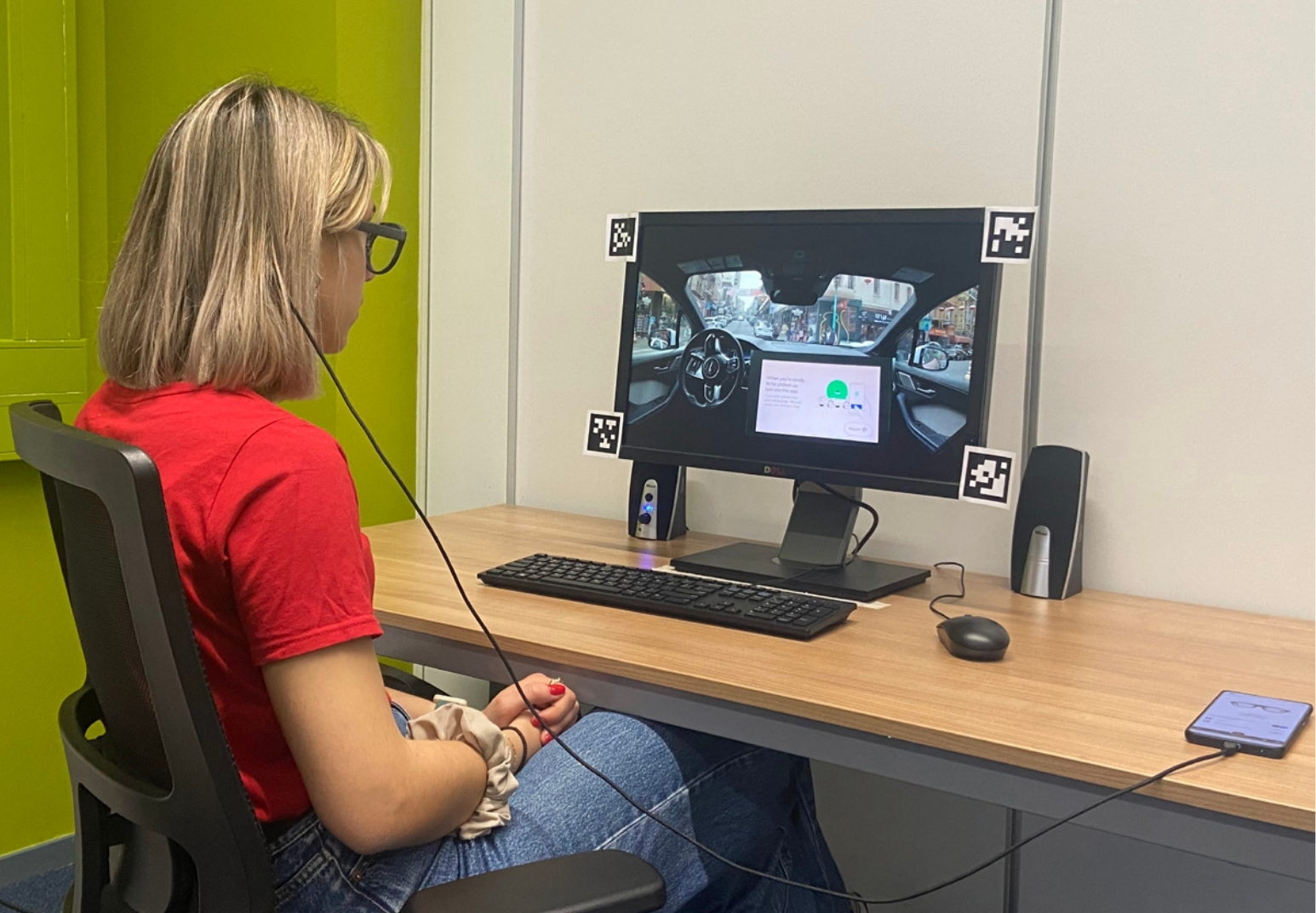}}
    \label{fig:participant}}%
    \caption{Both LLM- and human-generated mind maps were presented on a web-based platform (Fig. \ref{fig:platform}) for reviewing and editing design concepts during the lab study (Fig. \ref{fig:participant}).}%
    \label{fig:teaserfigure}
\end{teaserfigure}

\maketitle

\section{Introduction}
In recent years, design processes have increasingly incorporated diverse tools and methods. The accessibility of video recording devices has established video as a widely used medium in design. The approach, referred to as Video-Based Design (VBD), is employed to use videos to identify design challenges, draw inspiration, and develop effective solutions  \cite{jacob_design_2007, jacob_edit_2007, tatar_1989, vertelney_1989}. By leveraging ethnographic videos \cite{simonsen1997using}—recordings of human behaviours and interactions with products in situated environments—VBD provides designers with a rich, contextual understanding of user experiences. These videos serve as valuable tools to uncover latent issues and inspire innovative ideas \cite{jacob_design_2007}. However, ethnographic videos often contain fragmented elements, such as user interactions, environmental contexts, and personal narratives, which require substantial effort for designers to organize and distill into actionable insights. As Ylirisku and Buur emphasized, VBD demands that designers navigate large volumes of video content, extract the essence from the ethnographic footage, and systematically organize their findings into design decisions \cite{jacob_design_2007}. Traditionally, the organizing process in design relies on documentation techniques such as empathy mapping \cite{siricharoen2021using}, customer journey mapping \cite{rosenbaum2017create}, and experience mapping \cite{szabo2017user}, where designers use structured tables and visual frameworks to consolidate and interpret their insights cohesively. 

Mind maps, on the other hand, act as a powerful alternative for managing complex information in this context. Their intuitive structure often enables users to integrate diverse data types and support for cognitive processes such as memory recall and association in tasks. As Kedaj et al. \cite{kedaj2014effective} stated, mind maps provide a visual representation of hierarchical relationships and associations which simplify and organize multifaceted data especially in professional tasks. The flexible connections between concepts in mind maps allow users to classify ideas based on semantic connections and enhance retention \cite{zhang2010mind}. The use of mind maps for summarizing videos and enhancing professional tasks is increasingly prevalent. For instance, Siddarth et al. \cite{vimalaksha2019hierarchical} developed a framework to generate hierarchical mind maps from video lectures which simplifies lecture content into organized structures to aid learning. Similarly, the pipeline introduced by Zhao and Yang \cite{zhao2023promoting} uses mind maps to organize users' learnt knowledge with new concepts from tutorial videos to improve video-based learning. Additionally, Mammen et al. \cite{mammen2018beyond} demonstrated how mind maps can support qualitative data analysis by organizing video content using tools such as XMind into clear conceptual groupings for concepts. 

The rise of Large Language Models (LLMs) such as GPT-4 \cite{openai2023gpt4}, have demonstrated significant potential in organizing and synthesizing complex information \cite{bharathi2024analysis, zhao2023survey}. 
Prior research highlights the capabilities of LLMs in addressing challenges related to data integration, knowledge fusion, and information processing. For instance, Yin et al. explored how LLMs’ semantic reasoning abilities, combined with prompt instruction tuning, can enhance users’ decision-making by distilling critical information from heterogeneous data sources \cite{yin2023heterogeneous}. Similarly, Remadi et al. revealed that LLMs possess the capability to extract entities and resolve ambiguities within unstructured datasets \cite{remadi2024prompt}. Additionally, Zhang et al. introduced Video-LLaMA \cite{damonlpsg2023videollama}, an LLM framework capable of understanding content in videos. Their framework leverages an LLM model and enables the automatic generation of text-based descriptions from videos. Despite these contributions, prior research has not yet examined the application of LLMs to improve video understanding and information processing within the context of VBD.

To explore the utilization of LLMs in design, we investigate their potential to streamline VBD by reducing low-level human effort in video understanding and fostering efficiency in collecting design concepts and their relationships to one another (design information mapping). Specifically, we focus on the application of LLM-generated mind maps—structured, visual tools that represent hierarchical information and relationships—to assist designers in synthesizing insights from ethnographic videos and organizing their design ideas on a unified platform. We address the following research questions:
\begin{itemize}

\item[\textbf{RQ1:}] How are LLM-generated mind maps rated compared to human-generated ones in information mapping of video-based design (VBD)?
\item[\textbf{RQ2:}] In what ways do LLM-generated mind maps differ from human-generated mind maps in the effectiveness of practising in VBD workflows?
\item[\textbf{RQ3:}] What impact do LLM-generated mind maps have on the designers’ acceptance and perceived usefulness compared to human-generated mind maps in aiding VBD?
%\item[\textbf{RQ4:}] Which factors determine the effectiveness of LLM-generated mind maps in supporting designers’ decision-making during VBD?
\end{itemize}
To answer the questions, we conducted a controlled experimental study involving 28 designers from a university in scenario-based VBD exercises. We compared LLM-generated mind maps to human-generated ones to evaluate the performance of designers' information understanding and organizing processes in VBD. 
As results, all participants recognized the potential of LLM-generated mind maps to enhance efficiency and provide a starting point for VBD. Many appreciated their ability to automate the labour-intensive process of initial information capture which promoted ideation for higher-levelled tasks. Our findings also show that while LLM-generated mind maps offer significant advantages in automating data capture from ethnographic videos and providing a foundational structure, they face challenges in usability, organization, and decision-making support compared to human-generated maps. These insights showcase the great potential of LLMs in upscaling design processes while drastically reducing human effort. Specifically, compared to human-generated mind maps, LLM-generated maps are: \textbf{1)} \textit{More efficient in automating data capture but require more time for designers to edit and refinement.}
\textbf{2)} \textit{Less effective in organizing hierarchical structures.}
\textbf{3)} \textit{More demanding on cognitive load due to unstructured outputs, despite reducing initial manual effort.}
\textbf{4)} \textit{More reliant on trust, workflow integration, and human oversight to support effective decision-making.}

%All participants recognized the potential of LLM-generated mind maps to enhance efficiency and provide a starting point for VBD. Many appreciated their ability to automate the labor-intensive process of initial information capture which promoted ideation for higher-leveled tasks. However, participants also highlighted limitations in the structural coherence of LLM-generated maps. The absence of semantic hierarchical organization in LLM-generated mind maps often required more manual refinement in the practice. This highlights the need to develop LLM-supported tools that not only extract meta-information and automate processes but also organize the information in structured formats that align with designers' specific goals. The contributions of our work are as follows:

%\begin{itemize}
%\item We examine the comparative performance of LLM- and human-generated mind maps in aiding VBD efficiency and effectiveness.
%\item We identify structural and usability challenges in LLM-generated mind maps and their implications for cognitive load and workflow integration.
%\item We propose considerations for improving LLM tools to better support creative workflows by combining automation capabilities with design expertise.
%\end{itemize}

\section{Methodology}
We conducted a within-subject experimental study over three weeks to examine how LLM-generated mind maps compare to human-generated mind maps in supporting information mapping of VBD. The independent variable was the type of mind map, which included two conditions: human-generated and LLM-generated. We developed a workflow using GPT-4o \cite{openai2023gpt4} to generate mind maps from videos and integrated them into a web-based tool. The tool includes a video player and an interface that allows users to view and modify the mind maps. This section details participants, experimental procedure, and measurements of performance in information mapping of VBD.

\subsection{Pre-Study Preparation}
\label{sec:preparation}
\begin{figure}
  \centering
    \subfloat[\centering Autonomous car navigation]{{\includegraphics[width=.48\linewidth]{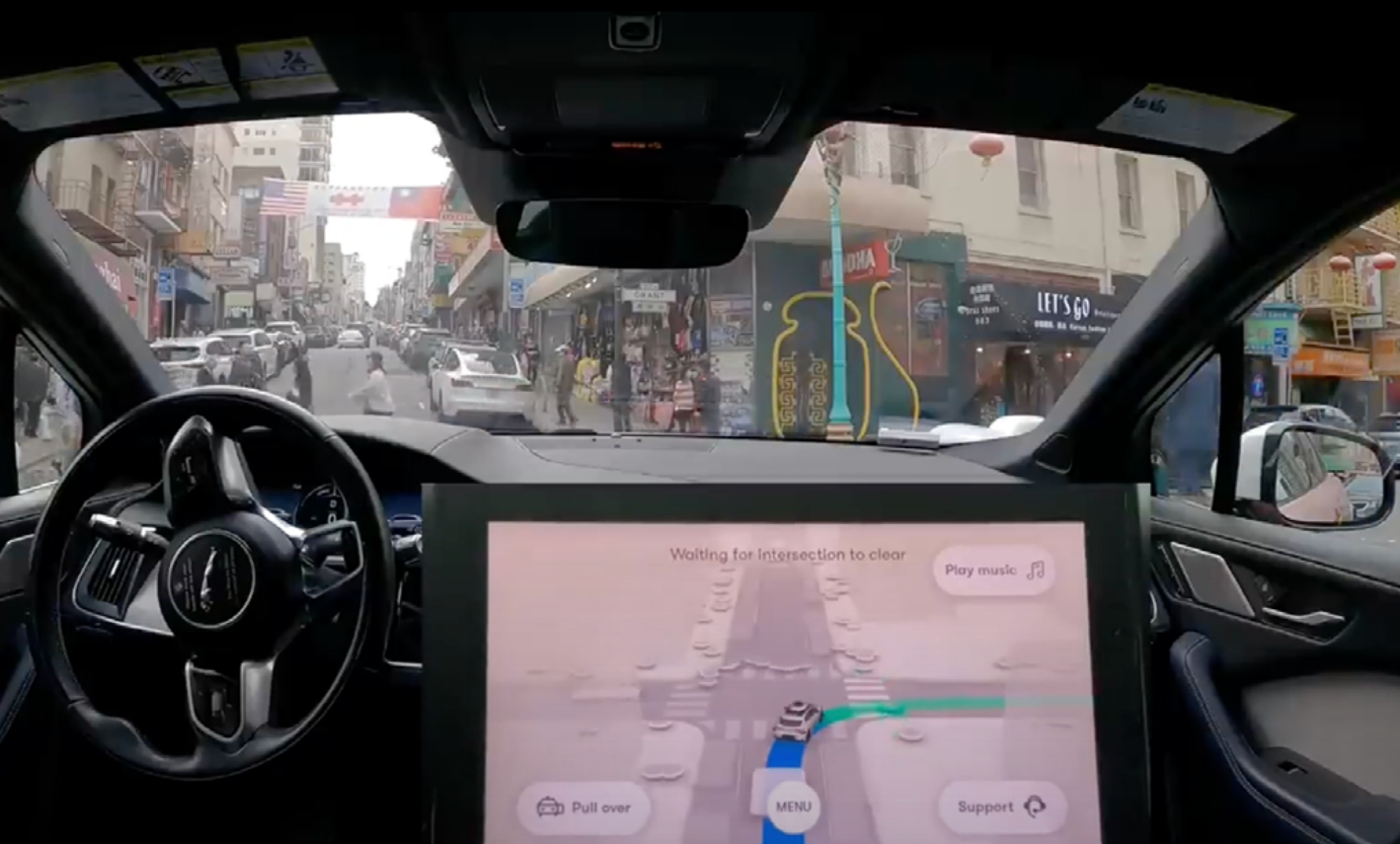}}
    \label{fig:car_context}}%
    \
    \subfloat[\centering Using accessibility features on a mobile phone]{{\includegraphics[width=.25\linewidth]{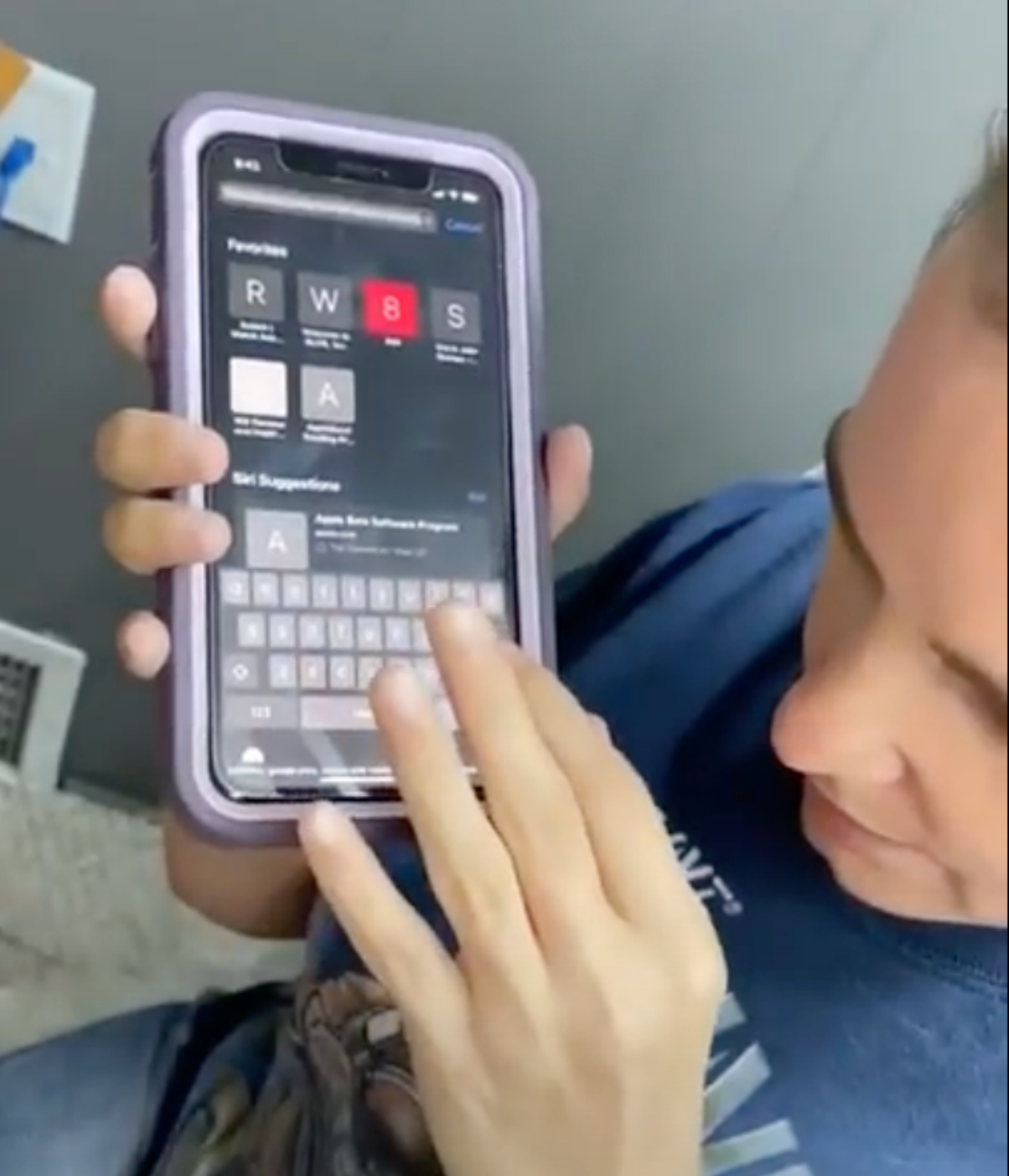}}
    \label{fig:phone_context}}%
    \\
    \caption{Screenshots for the two contexts used in the study: Fig. \ref{fig:car_context} shows a screenshot from a video about an autonomous taxi navigating a busy urban intersection with traffic signals, pedestrians, and other vehicles. Fig. \ref{fig:phone_context} shows a screenshot from a video showing a visually impaired user demonstrating accessibility tools on her phone, such as voiceover and screen reader functionalities, to navigate, type, and make a post on social media.}%
    \label{fig:video_contexts}
\end{figure}
We recruited 28 design students (9 females and 19 males) from our university. The participants had an average age of 25.8 years (SD = 1.9) and an average of 5.1 years of design experience (SD = 2.6). We asked participants to self-evaluate their experience with VBD (VBD-XP) and their perceived reliability of LLMs in daily tasks. Regarding LLM reliability, participants rated it as "high (daily)" (n=9), "moderate (weekly)" (n-9), and "rare (occasionally)" (n=10). Results indicate that 53.6\% of participants had high or moderate confidence in their VBD experience and 64.3\% relied on LLMs in their daily or weekly practices.

For design contexts, we selected two video scenarios each lasting 2 minutes and 20 seconds. These scenarios were carefully chosen based on professional designers' input to represent two primary categories of video content commonly encountered by designers: 1) repetitive and informative point-of-view recordings, and 2) user-product interactions. As indicated in Fig. \ref{fig:video_contexts}, the chosen contexts were an autonomous car navigating an urban environment (Fig. \ref{fig:car_context}) and a demonstration of a mobile phone accessibility features showed by a visually impaired user (fig. \ref{fig:phone_context}). Both videos were also standardized in bit rate to ensure consistency and fairness in participant evaluations. Based on the two selected videos, four mind maps were generated: one LLM-generated and one human-generated mind map for each video. 

To create the LLM-generated mind maps, we utilized \\ \texttt{blip2-opt-6.7b}\footnote{\url{https://huggingface.co/facebook/opt-6.7b} (last accessed: \today).}, a state-of-the-art Vision-Language Model (VLM), to transcribe the video content into textual descriptions. These descriptions were then processed using GPT-4 \cite{openai2023gpt4} with prompt fine-tuning to convert them into mind maps. In the fine-tuning process, the video descriptions along with their corresponding scenario topics, were incorporated into the prompts (see Supplementary Text \ref{prompt} in Appendix). We used prompts to instructed LLM to organize the video descriptions into a structured JSON format, which was then visualized as a mind map on the study platform. Additionally, for the human-generated mind maps, an independent designer was tasked with creating two mind maps for the two videos. The designer spent 10 minutes analysing each video to grasp its fundamental concepts. They then hand-drew a radial mind map on paper, which is a widely-used method in mind mapping to organize thoughts around a central theme \cite{buzan2006mind}. The designer spent an additional 20 minutes to digitally transfer the mind map into a JSON file which is similar to the LLM-generated ones. To keep pairwise comparison in the study, both the human- and LLM-generated mind maps were designed to maintain a similar number of topics, keywords and links between them as closely as possible.

We then developed a web-based platform (Fig. \ref{fig:platform}) consisting of two panels: a video player and a mind map editing panel. The video player on the left allowed participants to review and navigate the video content using controls such as play and pause. On the right side, the mind map editing panel displayed the mind maps visualized from JSON files. This panel enabled participants to actively edit the mind maps by moving nodes, modifying text, and adding or deleting links based on their understanding and interpretation of the video content and mind maps. Additionally, We included a countdown timer in the upper left corner to remind participants of the remaining time for their tasks.

\subsection{Measurements of Performance in Information Mapping}
\subsubsection{Mind Map Evaluation Rubrics} 
\label{sec: MMSR}
Due to time constraints in creating mind maps from scratch, our study used pre-created mind maps generated in advance using both LLM and human methods. Participants were asked to evaluate the mind maps from a designer's perspective and rate them using a holistic scoring approach based on the established Mind Map Scoring Rubric (MMSR) \cite{hua2019exploring}. The MMSR holistic scoring approach was chosen not only for its alignment with evaluators' perceptions of critical factors like accuracy and proficiency \cite{yune2018holistic} but also because it demonstrates higher inter-rater reliability for consistency compared to other qualitative rubrics \cite{tomas2019modeling}. In our study, we measured four key variables based on MMSR: 1) Identification of triggers (recognizing key concepts in the problem), 2) Development of concept links (exploring and expanding knowledge through valid connections), 3) Development of hierarchies (organizing concepts logically with core ideas at the center and specifics on the periphery), and 4) Identification of cross-links and relationship links (showing meaningful connections between different concepts and within a concept).

\subsubsection{Behaviour and Working Memory}
\label{sec: UTAUT2}
We adopted the questionnaire of Unified Theory of Acceptance and Use of Technology 2 (UTAUT2) \cite{tamilmani2021extended} to understand participant behaviour in mapping information. UTAUT2 measures participants’ willingness to integrate and use the mind maps on the study platform in their design workflows \cite{tamilmani2021extended}. The questionnaire assessed responses across eight categories: Performance Expectancy (perceived benefits), Effort Expectancy (ease of use), Social Influence (impact of others’ opinions), Facilitating Conditions (availability of resources and support), behavioural Intention (intent to use), Hedonic Motivation (enjoyment), Price Value (cost-effectiveness), and Habit (routine use). In addition, as working memory is critical for information processing and decision-making in advanced tasks \cite{paas2003cognitive}, we also assessed participants' cognitive load using NASA-TLX questionnaire \cite{hart1986nasa}. 
As an objective-based measurement of cognitive load \cite{ahlstrom2006using}, eye-tracking glasses (see Fig. \ref{fig:participant}) were also used to capture participants' eye movements, including saccades (rapid movements between focus points), fixations (sustained focus on a single point), pupil sizes (diameter of pupils), and blinks (rapid eyelid closures).

\subsection{Experimental Procedure}
\begin{table*}[h]
\centering
\caption{Overview of the experimental procedure with tasks performed by participants, and the measurements collected from preparation, main session and post-session of the study.}
\begin{tabular}{c c c c}
\toprule
\textbf{Process} & \textbf{Task} & \textbf{Measurements} \\
\midrule
\multirow{3}{*}{{Preparation}} & Signing consent forms & validity check\\ 
& Providing demographic information & validity check\\
& Having tutorials on study steps and tool usage & Inquiry and validity check\\
\midrule
\multirow{5}{*}{{Main Session}} & Watching video A & Eye-tracking\\
& Reviewing mind map A & Eye-tracking\\
& Evaluating mind map A & MMSR \cite{hua2019exploring}\\
& Modifying mind map A & Eye-tracking and Interaction logs\\
& Repeating tasks above with video B and mind map B & Measurements in task A\\
\midrule
\multirow{2}{*}{{Post-Session}} & Completing Post-Task Surveys & NASA-TLX \cite{hart1986nasa} and UTAUT2 \cite{tamilmani2021extended}\\
& Having an interview (10 min) & Qualitative feedback on mind maps' reasoning \\
\bottomrule
\end{tabular}
\label{tab:Experimental Procedure}
\Description{}
\end{table*}

As showed in Table \ref{tab:Experimental Procedure}, the experiment consisted of three phases: preparation, the main session, and post-session surveys and interview. Participants began by signing consent forms and provided demographic information. A brief tutorial introduced participants to the study's objectives, procedures, and the equipments used for the tasks. Participants were then assigned two tasks (A and B), which involved reviewing the two videos and editing the corresponding LLM- or human-generated mind maps. The order was predetermined using a randomized counterbalanced scheme. In the main session, participants began with Task A, where they first watched the assigned video (either context of autonomous car navigation or mobile phone accessibility features in Fig. \ref{fig:video_contexts}). They then reviewed the corresponding mind map A (either human- or LLM-generated) for 3 minutes. Participants then instructed to use the MMSR rubrics \cite{hua2019exploring} to evaluate mind map A on the four categories mentioned in Section \ref{sec: MMSR} on a scale from 1 to 100. Afterward, participants were given 10 minutes to modify mind map A using the web-based platform (see Section \ref{sec:preparation}). This process was repeated for Task B, where participants watched the second video (video B), reviewed, evaluated and edited the corresponding mind map B. Each task lasted approximately 20 minutes. By the end of the session, each participant was exposed to both versions of the mind maps (human- and LLM-generated) across the two video contexts. In the post-session, participants first completed the NASA-TLX questionnaire, followed by UTAUT2, and then interviewed about their experience in the study. %The follow-up discussion aimed to explore their reasoning and any distinguishing characteristics they observed. Participants were then thanked for their time and received a gift card as compensation. 

\section{Results \& Discussion}
We present the results of our experiment, which investigated the performance of information mapping using two methods across three dimensions: rating in effectiveness (RQ1), effectiveness in practice (RQ2), and use experience (RQ3). Statistical significance was determined using paired-samples t-tests and Wilcoxon signed-rank tests. Post-session interviews were analysed using thematic analysis \cite{ayre2022research} in ATLAS.ti to complement the quantitative results.

\subsection{LLM-Generated Mind Maps Are Effective for Concept Linking but Lag in Hierarchical Organization (RQ1)}
We first analysed how participants evaluated LLM- and human-generated mind maps using the four categories from the MMSR described in Section \ref{sec: MMSR}. No significant differences were found between LLM- and human-generated mind maps in three of the four categories: \textbf{identification of triggers} (Wilcoxon, Z = -0.054, p = 0.957), \textbf{development of concept links} (paired-samples t, t(27) = 0.478, p = 0.637), and \textbf{identification of cross-links and relationship links} (Wilcoxon, Z = -1.058, p = 0.290). It indicates that LLM-generated mind maps effectively identify key concepts from VBD videos, expand relevant knowledge, and establish meaningful connections. In this case, LLMs can generate mind maps with comparable information extraction and linkage to those created by professional designers, while potentially requiring less effort and reducing fatigue. However, a significant difference was observed in the category of the \textbf{development of hierarchies}, which measures the ability to logically organize and categorize concepts in VBD videos' information mapping. Human-generated mind maps scored significantly higher than LLM-generated ones (paired-samples t, t(27) = 2.456, p = 0.021). On average, human-generated mind maps scored 11.35\% higher in this category (LLM-generated: 50.79 ± 23.87 points, human-generated: 62.14 ± 23.99 points). While LLM-generated mind maps perform well in capturing and linking concepts, they fall short in logically structuring information with core and relevant ideas. From our interviews, a similar pattern was observed. Participants  (P4, P5, P11, P16, P18-21) noted that while LLM-generated mind maps enhance efficiency by providing a solid starting point with some connections between topics, human-generated information mapping still advantages in organizing clear categorization and deeper analysis on information. %Furthermore, human-generated mind maps were frequently praised for their clearer categorization and logical structure, with P6 noting, “[P6] It was more categorized, and the problem was identified in a very clear way.” While LLM-generated mind maps enhance efficiency by providing a solid starting point with some connections between topics, human-generated information mapping still advantages in organizing clear categorization and deeper analysis on information. 

\subsection{LLM-Generated Mind Maps Save Time but Require More Visual Effort and Trust-Building (RQ2)}
A Wilcoxon signed-rank test revealed that the time taken to analyse LLM- and human-generated mind maps in the study had no statistical significance (Z = -1.548, p = .122). However, a closer analysis showed that participants spent an additional 1.92 minutes on average editing LLM-generated mind maps compared to human-generated ones across the two design contexts (paired-samples t, t(28) = -2.278, p < .015; LLM-generated: 7.39 ± 3.58 min, human-generated: 5.47 ± 2.95 min). Furthermore, as cognitive load plays a significant role in influencing the effectiveness of information processing during design tasks \cite{ball2000putting, bartram2021untidy}, we employed NASA-TLX questionnaire to measure participants' working memory. A paired-samples t-test revealed no significant difference in self-reported cognitive load between the two conditions (t(27) = -1.291, p = .208). This indicates that participants perceived similar cognitive demands for both LLM- and human-generated mind maps. In addition to self-reported cognitive workload, we also used eye-tracking glasses to record participants' eye movements while reviewing and editing the mind maps. Paired-samples t-tests revealed that no significant differences found in eye fixation duration (t(27) = -.045, p = .964), blink duration (t(27) = .999, p = .327), or changes in pupil diameter (t(27) = -.870, p = .196) between the two conditions. However, a significant difference was observed in eye saccade duration, with participants spending 3.92\% more time sweeping through LLM-generated mind maps compared to human-generated ones (Wilcoxon, Z = -2.482, p = .013; LLM-generated: M = 60.96, SD = 21.76 ms; human-generated: M = 58.57, SD = 15.84 ms). Additionally, participants scanned the LLM-generated maps faster than the human-generated ones. A paired-samples t-test revealed that participants' eye movement speed was 4.46\% higher in the LLM-generated condition compared to the human-generated condition (t(27) = 2.113, p = .004; LLM-generated: M = 3674.97, SD = 561.03 px/s; human-generated: M = 3511.02, SD = 641.91 px/s). Participants were likely moving their gaze more quickly to navigate the elements in the LLM-generated maps. Together, these findings suggest that while the overall cognitive load was comparable across conditions, LLM-generated mind maps required more visual effort and faster scanning to interpret their contents effectively.

The later interviews confirmed the efficiency of both mind map types in formalizing information for VBD. P6 noted the time-saving aspect of LLM-generated mind maps: “[P6] One click and you get a mind map; it’s valuable but needs better structure.” Conversely, P25 appreciated human-generated mind maps for reducing the effort of video analysis: “[P25] It saves time and avoids burnout.” Some participants (P1, P16, and P18) highlighted the redundant keywords in LLM-generated mind maps but acknowledged the flexibility to ignore irrelevant elements. However, our further qualitative results revealed that the lack of trust from participants in LLM-generated mind maps significantly influenced their editing time. This resulted in having more time for additional visual and logical verification in the editing process. P25 stated that they did not "trust" the LLM-generated outputs, while P17 emphasized the need for "further adjustments" to make the LLM-generated mind maps more usable. As P5 mentioned, “[P5] You need to verify what’s in the video and the (LLM-generated) mind map. You shouldn’t just blindly accept it.” Similarly, P9 emphasized that information mapping “should involve human input” to ensure the accuracy and reliability of the LLM-supported tool for VBD tasks. 

\subsection{Human-Generated Mind Maps Are More Usable, While LLMs Stand Out in Efficiency (RQ3)}
We then analysed self-rated scores from the UTAUT2 framework (see Section \ref{sec: UTAUT2}) to measure participants’ acceptance and perceived usefulness of the two types of mind maps. In the categories of \textbf{social influence} (Wilcoxon, Z = -.460, p = .646), \textbf{facilitating conditions} (Wilcoxon, Z = -.577, p = .564), \textbf{hedonic motivation} (Wilcoxon, Z = -.408, p = .683), \textbf{price value} (Wilcoxon, Z = -.277, p = .782), and \textbf{habit} (Wilcoxon, Z = -1.112, p = .265), no significant differences were observed between LLM- and human-generated mind maps. However, paired-samples t-tests revealed significant differences between the two types of mind maps in the categories of \textbf{performance expectancy (PE)} (t(27) = 2.545, p = .017), \textbf{effort expectancy (EE)} (t(27) = 2.100, p = .045), and \textbf{behavioural intention (BI)} (t(27) = 2.464, p = .020). Participants reported higher scores for human-generated mind maps compared to LLM-generated ones across these three categories: PE (human-generated: 4.13 ± 0.53; LLM-generated: 3.88 ± 0.70), EE (human-generated: 3.88 ± 0.78; LLM-generated: 3.64 ± 0.72), and BI (human-generated: 3.87 ± 0.15; LLM-generated: 3.59 ± 0.16). From our interviews, participants frequently (n=11) noted that LLM-generated mind maps lacked the intuitive structure and customization designers expect. Participants also observed that LLM-generated mind maps lacked the contextual understanding of design concepts typically found in human-generated maps. On the other hand, some participants (n=6) acknowledged the LLM’s ability to capture extensive details and make gathering and organizing data "way more efficient" (P2). Additionally, we also found that ease of use and familiarity with the mind-mapping tool significantly influenced participants’ perceptions of LLM-generated maps. P11 noted some initial difficulties in interpreting the LLM-generated mind maps, saying, “[P11] I couldn’t get insights from the mind map because I didn’t know how to read the mind map (the connections between concepts).” This highlights the importance of intuitive design that facilitates quick comprehension and efficient navigation. However, training and continued use can mitigate these challenges: “[P13] It was more difficult at first, and then I found the value of it; it became easier for me.” Additionally, some participants (P23, P25-26) claimed that the flexibility in mind map editing and customization in concepts emerged as another critical factor for usability. The ability to modify and tailor mind maps not only enhances their relevance but also keeps designers actively engaged with the content and foster closer alignment with design goals.

%From our interviews, we also found that ease of use and familiarity with the mind-mapping tool significantly influenced participants’ perceptions of LLM-generated maps. P11 noted initial difficulty in interpreting the AI-generated mind maps, saying, “I couldn’t get insights from the mind map because I didn’t know how to read the mind map (add arrows).” This highlights the importance of intuitive design that facilitates quick comprehension and efficient navigation. However, training and continued use can mitigate these challenges: “[P13] It was more difficult at first, and then I found the value of it; it became easier for me.” Additionally, the flexibility in mind map editing and customization in keywords emerged as another critical factor for usability: “[P23] You can just change stuff, add things, say, ‘This is wrong. This is right,’ and that’s easier for the designer than starting from scratch.” This adaptability potentially allows designers to refine mind maps to meet their specific needs in a more efficient workflow. The ability to modify and tailor mind maps not only enhances their relevance but also keeps designers actively engaged with the content and foster closer alignment with design goals.

\section{Conclusion and Future Work}
This LBW provides early insights into the use of LLM-generated mind maps in supporting information mapping in VBD. Through a controlled experimental study with 28 designers, we evaluated the performance of LLM-generated mind maps compared to human-generated ones across self-rated scores, effectiveness and user acceptance. LLM-generated mind maps demonstrated advantages in effective collecting data captured from videos in VBD, consistent with previous studies that highlight the effectiveness of LLMs in processing complex information \cite{wen2023mindmap, huang2024vtimellm}. However, compared to the human-generated baseline, LLM-generated mind maps were often criticized for their lack of coherent hierarchies and contextual understanding, and  usually required additional manual refinement. Factors such as trust, transparency, and the ability to customize were also identified as critical to the acceptance of LLM-generated mind maps in VBD tasks.

To enhance the utility of LLM-generated mind maps in VBD's information mapping workflows, we will focus on improving structural and contextual alignment with design processes. We will start from refining prompt templates \cite{zamfirescu2023johnny} or developing better technical solutions for distilling LLM prompts \cite{li2023prompt} to enhance the hierarchical organization of design concepts and foster further customization and usability. Additionally, in our future studies, transparency should also be prioritized to build trust and improve usability. Implementing mechanisms such as detailed model reporting, clear explanations of LLMs' limitations, and visual feedback on reliability can help users better understand the capabilities and constraints of AI-generated outputs \cite{liao2023ai}. Finally, as highlighted in previous research \cite{graf2023chatgpt}, presenting LLMs' uncertainty using less precise but interpretable language could further enhance both the usability and trustworthiness of LLM-generated contents in future information processing tasks.

\bibliographystyle{ACM-Reference-Format}
\bibliography{main.bib}

\appendix
\begin{quote}
    \framebox{\parbox{\dimexpr\linewidth-2\fboxsep-2\fboxrule}{
\texttt{Output a JSON as plain text to describe the scenario. The JSON should contains nodes whose “label”
are related keywords from the topic and edges with “label” as relationships. Include unique ID, position,
size, shape for nodes; source, target for edges; and styles for both, with edge length reflecting semantic
relevance. Do not put the same IDs for edges. Depend on the contents, generate more than 20 but not
less than 30 nodes and edges, and generate 1 to 3 levels of branches as subtopics. Use different node and
edge colors and edge lengths to represent relationships. Avoid overlapping. Focus on objective interactions.
This is an example: “{
 “nodes”: [
 {“id”: “Node1”, “x”: 50, “y”: 50, “size”: [60, 60], “shape”: “circle”, “label”: “Apple”},{“id”: “Node2”, “x”: 200, “y”: 50, “size”: [60, 60], “shape”: “circle”, “label”: “iOS”}, {“id”: “Node3”, “x”: 350, “y”: 50, “size”: [60, 60], “shape”: “circle”, “label”: “App Store”}, ...]}”. Use the example as a template for generation but follow the rules above. This is the scenario: “...”. Here
is the transcript: “...”. }
}
}
\captionof{floatquote}{Instructional text displayed .}
\label{prompt}%
\end{quote}
\end{document}